\documentclass[manuscript]{aastex}
\usepackage{graphicx}
\usepackage{amsmath}
\usepackage{aas_macros}

\newcommand\beq{\begin{equation}}  
\newcommand\eeq{\end{equation}}  
\newcommand\D{\Delta}
\newcommand\sint{\int_{\theta,\phi \in S^2} \!\!\!\!\!\!\!\!\!\!\! d\theta d\phi\;}

\newlength{\figwidth}
\begin{document}
\title{Fast and Exact Spin-$s$ Spherical Harmonic Transforms}
\author{Kevin~M.~Huffenberger}
\affil{Department of Physics, University of Miami, 1320 Campo Sano Dr., Coral Gables, FL 33146}
\email{huffenbe@physics.miami.edu}

\and

\author{Benjamin~D.~Wandelt}
\affil{Department of Physics, University of Illinois at Urbana-Champaign, 1110 W Green Street, Urbana, IL 61801}
\affil{Institut d'Astrophysique de Paris, UMR 7095 CNRS--Universit{\'e} Pierre et Marie Curie, 98bis, bd Arago, F-75014 Paris, France}

\begin{abstract}
We demonstrate a fast spin-$s$ spherical harmonic transform
algorithm, which is flexible and exact for band-limited functions. In contrast to previous work, where spin transforms are computed independently, our algorithm permits the computation of several distinct spin transforms simultaneously. Specifically, only one
set of special functions is computed for transforms of quantities with
any spin, namely the Wigner $d$-matrices evaluated at $\pi/2$, which may be computed with efficient recursions.  For any spin the
computation scales as  ${\cal O}(L^3)$ where $L$ is the band-limit of the function.  Our publicly available numerical implementation permits very high accuracy at modest computational cost.  We discuss applications to the Cosmic Microwave Background (CMB) and gravitational lensing.
\end{abstract}


\section{Introduction}
Spin-weighted spherical harmonic functions generalize the standard spherical harmonics to allow the representation and decomposition of spin-weighted functions on the sphere.  
Introduced to study gravitational radiation and the electron's wavefunction in a magnetic monopole field \citep{newman:863,Wu1976365,dray:1030}, spin harmonics have several applications in astrophysics and cosmology.  In particular, they are useful for statistical studies of  orientation-dependent signals on the celestial sphere, such as cosmic shear measurements of lensed galaxies, cosmic microwave background (CMB) polarization, and gravitational lensing of polarized radiation, among others.

A spin-$s$ function on the sphere, which transforms under a local rotation by angle $\psi$ as $f \rightarrow e^{is\psi}f$, may be decomposed as
\beq
f(\theta,\phi) = \sum_{l=0}^{L} \sum_{m=-l}^{m=l}  \ {}_s a_{lm} \ {}_s Y_{l m}(\theta,\phi)
\eeq
using the set of spin spherical harmonics ${}_s Y_{l m}(\theta,\phi)$, complex-valued functions on the sphere.  Standard scalar spherical harmonics are represented by $s=0$; from these we obtain the spin harmonics using spin-raising and lowering operators \citep[e.g.][]{goldberg:2155}.  The decomposition depends on the orthogonality and completeness relationships
\begin{eqnarray}
 \int_{\theta,\phi \in S^2} \!\!\!\!\!\!\!\!\!\!\! d\theta d\phi\; {}_sY_{lm}(\theta,\phi)\ {}_s{Y}^*_{l'm'}(\theta,\phi)\ &=& \delta_{ll'} \delta_{mm'}, \\ \nonumber
\sum_{l m} {}_s Y_{l m}(\theta',\phi') {}_s Y^*_{l m}(\theta,\phi) &=& \delta(\phi'-\phi)\delta(\cos\theta'-\cos\theta).
\end{eqnarray}

For a band-limited function with ${}_s a_{lm} \neq 0$ only for $l \leq L$, the brute force computation of the transform requires  ${\cal O}(L^4)$ operations: roughly, there are $\sim L^2$ terms in the sum for each of the $\sim L^2$ points on the sphere required to Nyquist sample the function.  In this work, we describe a ``fast'' transform, which performs the analysis and synthesis for spin-$s$ functions in ${\cal O}(L^3)$ operations on an equiangular grid.

Several methods exist for spin transforms on the sphere, but our method offers some advantages.  The \citet{DriscollandHealy} algorithm for a scalar spherical harmonic transform scales very well, as ${\cal O}(L^2 \log^2 L)$, and relates the required Legendre transform to a fast Fourier transform (FFT) on a specific equiangular grid.  \citet{2007JCoPh.226.2359W} used that 
method and further relations between the scalar harmonics and the $s=\pm 2$ harmonics, implementing spin transforms at the same scaling.  The difficulty with these methods is that they require a large amount of memory to store precomputed special functions.  These precomputed functions require ${\cal O}(L^3)$ operations and storage, and so spoil the efficiency for computing a single transform, but need to be computed only once, in principle.  

Others have computed spin harmonic transforms on another pixelization of the sphere at ${\cal O}(L^3)$ scaling without precomputations \citep[e.g.~][]{2005ApJ...622..759G,2005PhRvD..71h3008L}.  However, these algorithms use recursions which depend on spin, so the individual spin transforms must be computed independently.  Our  ${\cal O}(L^3)$ algorithm alleviates this difficulty, since the same recursion is used for all transforms, and multiple distinct spin transforms can be computed in a single pass of a general code.  The approach in \citet{2008arXiv0807.4494M} is similar to ours, but differs in the details of the transforms, and suffers from numerical stability problems even at modest multipoles.  Our approach is numerically stable to high multipoles.

We present our algorithm in section \ref{sec:method}, describing forward and inverse transforms, and our publicly-available numerical implementation.  
In section \ref{sec:discussion} we discuss our results and conclude in \ref{sec:conclusions}.


\section{Method} \label{sec:method}
We exploit the well-known relationship \citep{goldberg:2155} between the spin-$s$ spherical
harmonics and the Wigner $d$-functions, which can be used to characterize rotations in harmonic space:
\beq
\ _s Y_{l m}(\theta,\phi)=(-1)^s \sqrt{\frac{2l+1}{4\pi} }
d^l_{m(-s)}(\theta) \exp({im\phi}).
\label{sYlmdlmm}
\eeq
The $d$-function is associated with an active right handed rotation about the $y$-axis by
$\theta$, a rotation which we can alternatively represent as a sequence of 
five
 rotations: 
(1) right by $\pi/2$ about the $z$-axis, 
(2) right by $\pi/2$ about the $y$-axis,  
(3) right by $\theta$ about the $z$-axis, 
(4) left by $\pi/2$ about the $y$-axis, 
and (5) left by $\pi/2$ about the $z$-axis.
Defining
\beq
\D^l_{mm'}\equiv d^l_{mm'}(\pi/2),
\eeq
this allows use to rewrite the Wigner $d$-functions as \citep{1996JGeod..70..383R}:
\beq
d^l_{MM'}(\theta)= i^{M-M'}\sum_{m=-l}^{m=+l} \D^l_{mM} \exp({-im\theta})
\D^l_{mM'} 
\label{factorize}
\eeq
This seemingly complicated procedure has two advantages: (1) all
rotations about the $y$-axis can be represented in terms of the Wigner 
$d$-functions evaluated only at $\pi/2$; and (2) sums over
exponentials can be done efficiently with Fourier transforms.
This particular factorization is a special case of the more general
representation of convolution on the sphere presented in
\citet{2001PhRvD..63l3002W}. 

The $\Delta$-matrices can be computed quickly and easily using recursion relations \citep{1996JGeod..70..383R,Trapani:cc5006}.  Several symmetries allow us to improve the computational efficiency, so only $\sim 1/8$ of the matrix entries need to be computed (appendix \ref{app:wigner_symm}).

\subsection{Forward  transform}
The forward (direct or analysis) spin-$s$ spherical harmonic transform of a spin-$s$ function
$f(\theta,\phi)$ is defined as
\beq
_s a_{lm} = \sint \ _s Y^\ast_{l m}(\theta,\phi)
f(\theta,\phi)\sin{\theta}.
\eeq
Using the relationship to Wigner $d$-functions, the transform becomes
\begin{eqnarray} \nonumber
\ _s a_{lm}&=&\sint (-1)^s \sqrt{\frac{2l+1}{4\pi}}
d^l_{m(-s)}(\theta) \exp({-im\phi}) f(\theta,\phi)\sin{\theta}\\ \nonumber
&=& (-1)^s \sqrt{\frac{2l+1}{4\pi}} 
 i^{m+s} \sum_{m'=-l}^{m'=+l} \D^l_{m'm}  \D^l_{m'(-s)} \sint \exp({-im'\theta})
 \exp({-im\phi}) \sin{\theta} f(\theta,\phi) \\
&=& (-1)^s \sqrt{\frac{2l+1}{4\pi}} 
 i^{m+s} \sum_{m'=-l}^{m'=+l} \D^l_{m'm}  \D^l_{m'(-s)} I_{m'm},
\end{eqnarray}
where we have ignored, for the moment, the details of evaluating the
integral,
\beq
I_{m'm}\equiv\sint  \exp({-im'\theta})
 \exp({-im\phi}) \sin{\theta} f(\theta,\phi). \label{eqn:Imm}
\eeq
This is a
quadrature problem in the form of a 2D Fourier integral.  In appendix~\ref{quadrature} we show that for a band-limited function these $(2L+1)^2$
integrals can be performed exactly in
${\cal O}(L^2 \log L)$ operations by simply evaluating a two-dimensional FFT of a modified form of the function multiplied by an easily computed set of quadrature weights.

Since each component of $\D^l_{mm'}$ can be computed using ${\cal O}(1)$ floating point evaluations \citep[\textit{e.g.}][]{1996JGeod..70..383R,Trapani:cc5006}, we
have a formula that evaluates the spherical 
harmonic coefficients of the general spin $s$ transform by evaluating a
single sum over $m'$ for each $l$ and $m$, with asymptotic scaling ${\cal O}(L^3)$.



A symmetry on the first azimuth index of the $\Delta$ allows a more efficient representation of the sum, which cuts the computation time in half:
\begin{equation}
 \sum_{m'=-l}^{m'=+l} \D^l_{m'm}  \D^l_{m'(-s)} I_{m'm} =  \sum_{m'=0}^{m'=+l} \D^l_{m'm}  \D^l_{m'(-s)} J_{m'm}
\end{equation}
where
\begin{equation}
J_{m'm} = \left\{ \begin{array}{ll}
I_{0m} & \mbox{if $m'=0$} \\
I_{m'm} + (-1)^{m+s} I_{(-m')m} & \mbox{if $m' > 0$ } 
\end{array} \right.
\end{equation}
and $J_{m'm}$  is undefined for $m' < 0$.

The symmetry on the second azimuth index of the $\Delta$ allows us to rewrite the expression so that it requires only values from the $\Delta$ matrices in the non-negative quadrant, which allows more efficient use of the recursion and allows reuse of the $ \D^l_{m'm}  \D^l_{m'(-s)} $ product for another modest efficiency gain.


If $f$ is real, the Fourier content is halved, and $I_{(-m')(-m)} = I^*_{m'm}$, implying $J_{0(-m)} = J^*_{0m}$ and  $J_{m'(-m)} = (-1)^{m+s}J^*_{m'm}$ (for $m'>0$).  We use this relationship to cut the computation time by another factor of two.  Additionally, realness implies ${}_{-s} a_{l(-m)} = {}_{s} a_{lm}^* $, providing some transforms for free.

\subsection{Inverse Transform}
The backward (inverse or synthesis) spherical harmonic transform maps the $\ _s
a_{lm}$, $s \le l\le L$, into a function on the sphere 
\beq
f(\theta,\phi)=\sum_{lm}\ _s a_{lm} \ _s Y_{l m}(\theta,\phi).
\eeq
Since the inverse transform does not involve an integral, issues
of quadrature accuracy do not arise.  We can again use the Wigner function relation to write this as
\begin{eqnarray} \nonumber
f(\theta,\phi)&=& (-1)^s \sum_{m m'} \exp({im'\theta})  \exp({im\phi}) i^{m+s}  \sum_{l} \sqrt{\frac{2l+1}{4\pi}
} \D^l_{(-m')(-s)} \D^l_{(-m')m}  \ _s  a_{lm}\\
&=& \sum_{m m'} \exp({im'\theta})  \exp({im\phi})\  G_{m'm}
\end{eqnarray}
The $m$ and $m'$ sums can be computed using FFTs, and are sub-dominant to the scaling, using ${\cal O}(L^2\log L)$ operations.  The FFTs will produce
$f(\theta,\phi)$ as a 
regularly sampled function on a 2-torus.  Only half of this torus is of
interest and the $\theta > \pi$ portion can be ignored. An appendix to \citet{2008arXiv0811.1677B} also noted this algorithm for the inverse transform (but does not address the forward transform).

Computation of 
\beq
G_{m'm} =  (-1)^s  i^{m+s}  \sum_{l} \sqrt{\frac{2l+1}{4\pi}
} \D^l_{(-m')(-s)} \D^l_{(-m')m}  \ _s  a_{lm}
\label{eqn:Gmm}
\eeq
 again scales as ${\cal O}(L^3)$.
Mirror symmetries of the Wigner matrices again allow the expression to be rewritten using only the non-negative quadrant.
The mirror symmetry on the first azimuthal index of $\Delta$ leads to
\beq
G_{m'm} = (-1)^{m+s} G_{(-m')m}
\eeq
which cuts the computation time in half. 
For a transform to a real-valued function, the additional symmetry
$G_{(-m')(-m)} = G_{m'm}^* $
can again cut the computation time by two.

\subsection{Numerical implementation}\label{sec:implementation}

We implement these transforms for double precision complex-valued functions on the sphere.  A C library is available from the authors\footnote{http://www.physics.miami.edu/$\sim$huffenbe/research/spinsfast}.  In this implementation, the equiangular (Equidistant Cylindrical Projection or ECP) pixelization is described by parameters $N_\phi$ and $N_\theta$, describing the number of pixels in each direction.  The first pixel column is centered at $\phi = 0$. The first and last rows each contain redundant pixels at the poles, $\theta = 0$ and $\theta = \pi$, respectively.  The FFT portions of the computation use the FFTW library\footnote{http://www.fftw.org} and are most efficient when the dimensions of the $\theta$-extended map (appendix~\ref{quadrature}) are are powers of two.  The dimensions of the extended map in our implementation are $N_\phi$ and $2 (N_\theta - 1)$.  To support a function with limited bandwidth $l \leq L$ requires $N_\phi, N_\theta \geq 2 L+1$.

\begin{figure}
\begin{center}
\includegraphics[width=\figwidth]{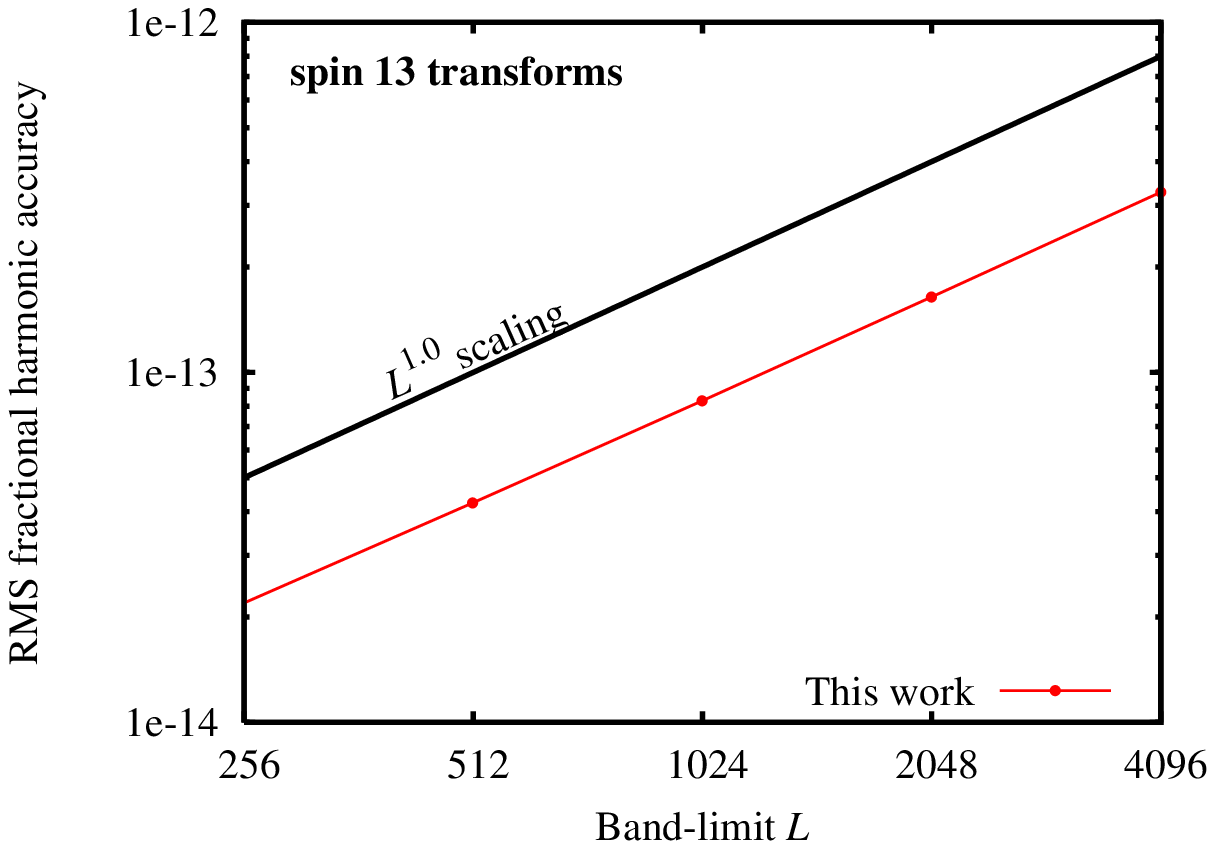}
\end{center}
\caption{Accuracy of a backward-forward transform pair for Gaussian random white noise.  HEALPix transform pairs (without iteration) yield an accuracy $\sim 10^{-3}$ in the same test.}
\label{fig:accuracy}
\end{figure}

We verified the accuracy of our method in two ways.  
First, for selected low multipoles, we checked our transforms against the analytic expressions for single spin spherical harmonics from \citet{goldberg:2155}.  
Second, we showed that our transforms are extremely accurate by comparing the spherical harmonic coefficients from a backward-forward transform pair to the original coefficients (Fig.~\ref{fig:accuracy}), up to a band-limit of $L=4096$.  We plot the rms of the relative difference between the original ($_sa_{lm}$) and recovered  ($_s\hat{a}_{lm}$) harmonic coefficients, $|_sa_{lm} - _s\hat{a}_{lm}|/|_sa_{lm}|$, for Gaussian random white noise, showing typical errors of $\sim 10^{-13}$ which increase roughly proportional to the band limit.  Comparison of the maximum relative errors shows that these transforms are 3-4 orders of magnitude more accurate than the (already very accurate) $s=0,\pm2$ transforms from \citet{2007JCoPh.226.2359W} onto a similar grid.   

HEALPix\footnote{http://healpix.jpl.nasa.gov} \citep{2005ApJ...622..759G}, the very useful numerical grid and associated software suite for representing and manipulating functions on the sphere, has found widespread use in the astronomy and astrophysics community, particularly for CMB work.  Under the same test, HEALPix transforms, which are approximate, show typical errors of $\sim 10^{-3}$ (for resolution parameter $N_{\rm side} = L/2$, a commonly used value).  HEALPix provides an option to iterate the transform to converge to somewhat higher accuracy, if needed.

\begin{figure}
\begin{center}
\includegraphics[width=\figwidth]{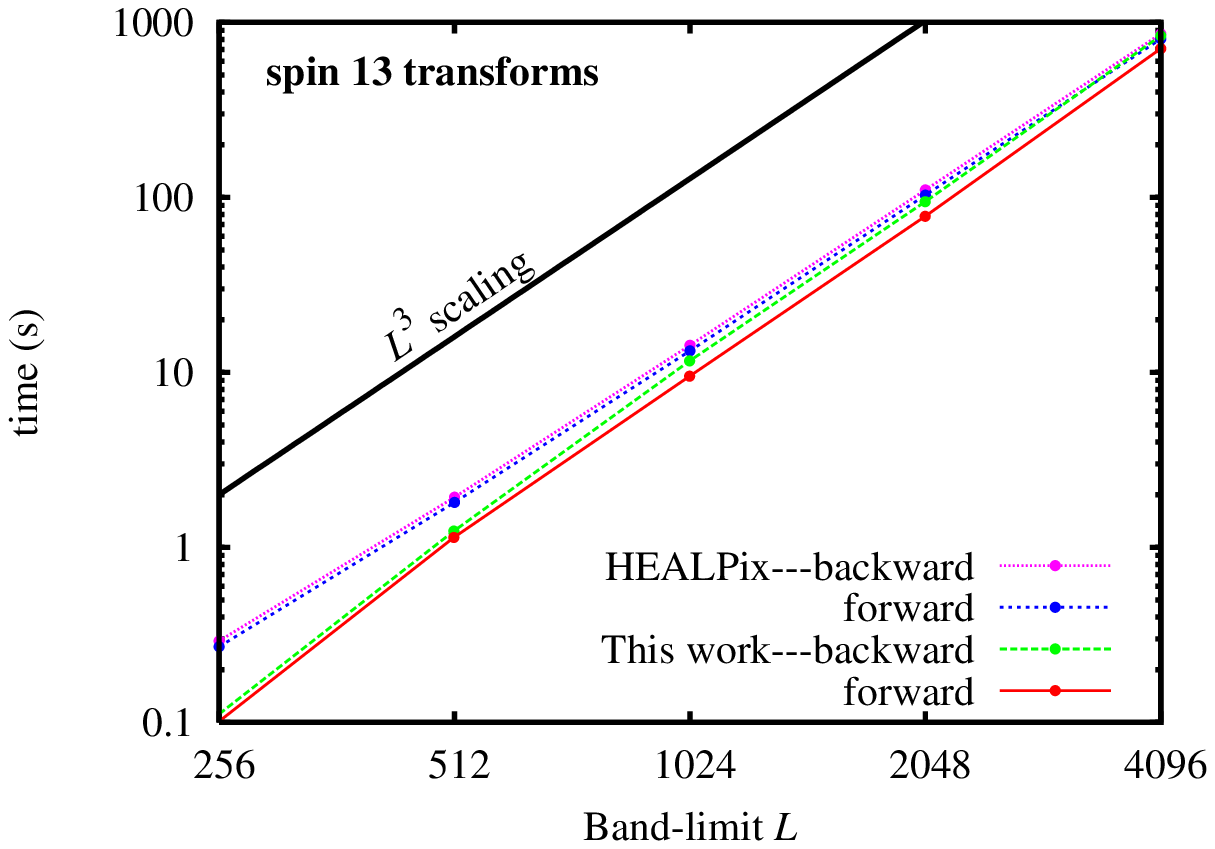}
\end{center}
\caption{Speed comparisons for spin-13 transforms of random spherical harmonic coefficients.  HEALPix routines use parameter $N_{\rm side} = L/2$.  The fast spin transform was run with the minimum pixel size necessary to support the bandwidth, as described in the text.} \label{fig:speed}
\end{figure}

Timing tests were performed on a single 2.83 GHz Xeon processor.
Fig.~\ref{fig:speed} shows that the scaling of this implementation is approximately $L^3$.  The speed is comparable to or slightly faster than the similar arbitrary-spin transform computed by the HEALpix suite (version 2.11).  The comparison includes the times for the recursive computation of special functions---associated Legendre functions for HEALPix and Wigner $d$-functions for our algorithm---and both may be accelerated by precomputing these quantities (at ${\cal O}(L^3)$ memory cost).

For the HEALPix comparison, again we use $N_{\rm side} = L/2$, which means the HEALPix transforms use $3 L^2$ pixels.  In the same comparison our FFT code uses about $4L^2$ pixels, the minimum number to fully support the band-limited function.  If both the band-limit and the number of pixel were set the same, our algorithm would fare slightly better than shown here.

\begin{figure}
\begin{center}
\includegraphics[width=\figwidth]{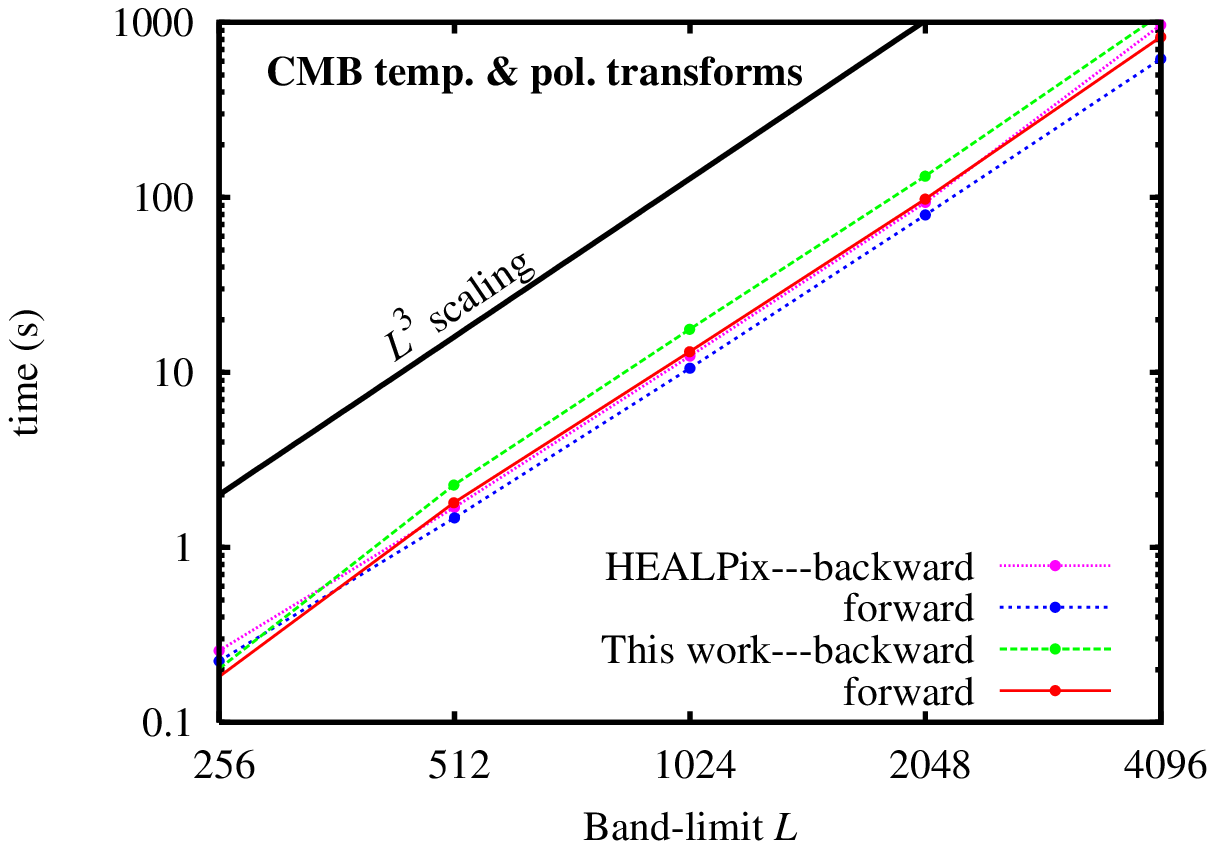}
\end{center}
\caption{Similar to figure \ref{fig:speed}, but for the harmonic transforms of temperature and polarization Stokes parameters, commonly used in the analysis of CMB data.} \label{fig:speed_iqu}
\end{figure}

We also implemented transforms for temperature and polarization Stokes parameters $(T,Q,U)$ which commonly occur in CMB analysis \citep[see][noting the differing phase convention for $Y_{lm}$]{1997PhRvD..55.1830Z}:
\begin{eqnarray}
  T(\theta,\phi) &=& \sum_{lm} a_{T,lm}\  Y_{lm}(\theta,\phi) \nonumber \\
  (Q \pm iU)(\theta,\phi) &=& \sum_{lm} a_{\pm 2,lm}\  {}_{\pm 2}Y_{lm}(\theta,\phi)
\end{eqnarray}
These quantities permit some additional symmetries: $T$ is real-valued and 
$a_{-2,lm} = (-1)^m a^{*}_{+2,l(-m)}$.  In Fig.~\ref{fig:speed_iqu} we compare an implementation of our algorithm to the temperature and polarization HEALPix transforms, which are more highly optimized than those for a general spin.  Again the speed is comparable, this time with HEALPix slightly faster.

In our implementation, roughly half the computation time is spent computing the Wigner $\Delta$-functions.  These use recursive algorithms, and may be computed during the course of the transform, or precomputed at memory cost of $\sim L^3/6$ floating point numbers.   The method for computing the $\Delta$ functions may be selected at runtime.  We have implemented algorithms for the Wigner matrices from \citet{1996JGeod..70..383R} and \citet{Trapani:cc5006} (represented internally at quadruple precision).  In our implementation, the \citet{Trapani:cc5006} recursion is faster and more accurate to our maximum test bandlimit $L = 4096$, but this recursion eventually goes unstable at very high multipoles.  If they are used at double precision, instead of quadruple, the transforms proceed about 10 percent faster, but go unstable between $L=2048$ and $L=4096$.

 The transform times listed in \citet{2007JCoPh.226.2359W} for $L \leq 1024$ are comparable to ours, although these do not include the significant precomputation time.  Methods based on \citet{DriscollandHealy}, such as \citet{2007JCoPh.226.2359W}, should be much faster at higher band-limits because of the improved scaling, but the memory needs become impractical at present, requiring $\sim 77$ GB of memory for $L=4096$ (scaling as $L^3$ from the \citet{2007JCoPh.226.2359W} memory value for their maximum band-limit, $L=1024$).

\begin{figure}
\begin{center}
\includegraphics[width=\figwidth]{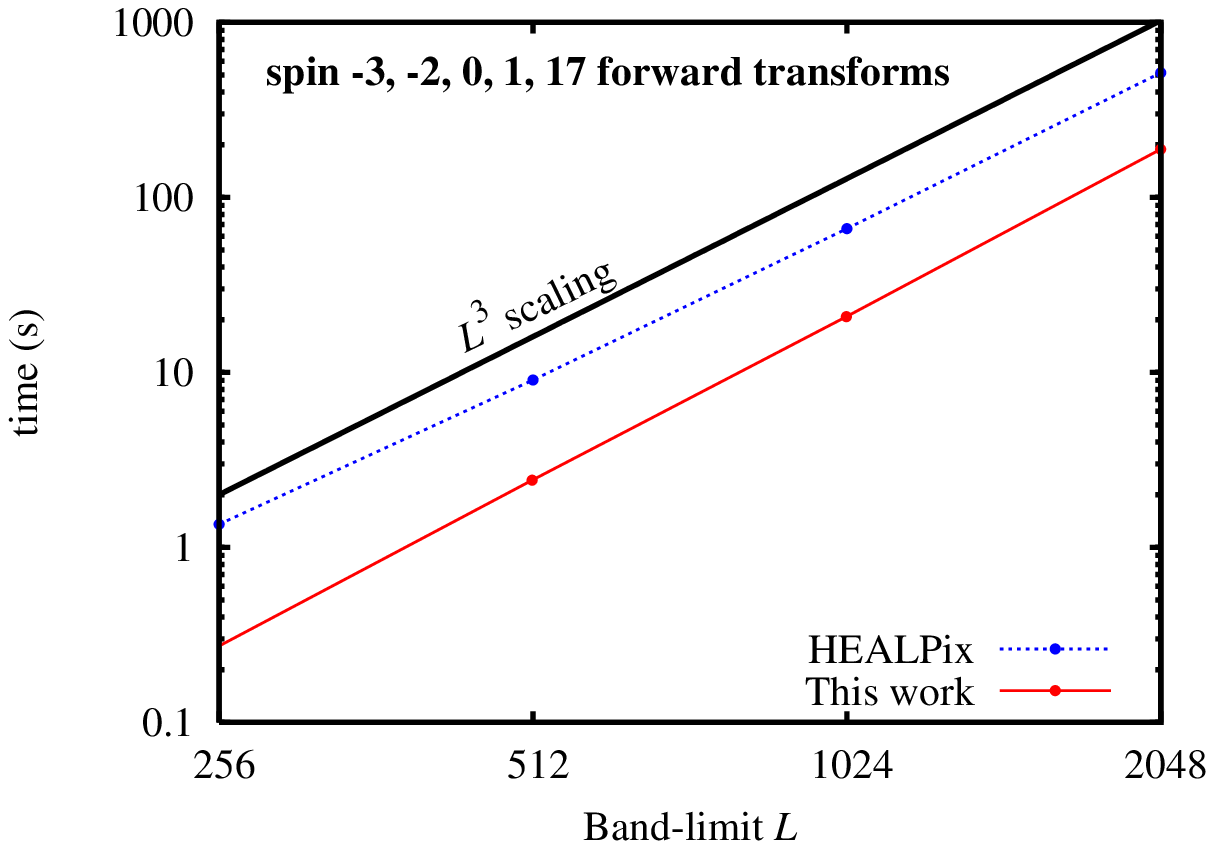}
\end{center}
\caption{Computing five different spin transforms (forward transform only).  The HEALPix transforms are computed successively.  Our algorithm reuses the same Wigner $d$-function recursion for all transforms, leading to a substantial speed-up.}
\label{fig:speed-multi}
\end{figure}

Another strength of our approach is that we can re-use the same Wigner matrix to compute simultaneously several transforms with differing spins.  Because the recursions are a significant portion of the computation, this can be much more efficient than running the transforms independently.  Fig.~\ref{fig:speed-multi} shows the computation time for five spin transforms computed together.  Our code can reuse the Wigner matrix recursion for each transform, in contrast to the HEALPix transforms, which must recompute the required recursions for each spin order.  For a large number of transforms, such that the Wigner recursion is sub-dominant, our transforms proceed about twice as fast as if they were computed individually.



\section{Discussion} \label{sec:discussion}
Our algorithm is convenient because it implements all spin-$s$
transforms in a single routine, and is particularly
efficient if transforms at several spin orders have to be computed at
the same time. Because of the representation of the spherical harmonic
coefficients in terms of Wigner $d$-functions at a single argument, no
additional computations of special functions are necessary if we would
like to do several spin-s transform at once. For CMB
 work we have to
compute spin 0 and $\pm 2$ transforms for polarization, as well as spin 1 and spin 3 transforms to simulate lensing. Our generalization therefore presents economies of scale. 

The time and memory requirements for the ${\cal O}(L^3)$ portion of the calculation depends only on the band limit and not the number of pixels.  Unlike HEALPix, oversampling the function in real space costs very little, only a 2-dimensional FFT, so synthesizing or analyzing a map with a modest band limit at high resolution---zero padding the harmonic object---is relatively painless, and is typically bounded by the memory available, not the computation time.

This has immediate application to the simulation of CMB lensing.  Unlensed CMB maps are simulated as isotropic fields by taking the spherical harmonic transform of Gaussian random harmonic coefficients.  The gravitational potential deflects the path of light rays from the CMB's last scattering surface.  Hence, a pixel's light will not originate from that, or any other, pixel's center, complicating the evaluation of the spherical harmonics.  A typical strategy is to highly oversample the unlensed map, then based on the deflection field remap pixels \citep{2005PhRvD..71h3008L} or use an interpolation scheme \citep{2004PhRvD..70j3501H,2008ApJ...682....1D,2008arXiv0811.1677B}.  The algorithm described here makes this easier at high resolution.  In particular, \citet{2008arXiv0811.1677B} uses our same method for the backward transform, and showed these interpolations can be carried out with very accurately.  Also, the harmonic object $C_{n,m}$ constructed as part of the interpolation in \citet{2004PhRvD..70j3501H} and \citet{2008ApJ...682....1D} by a Legendre transform followed by an FFT is equivalent our $G_{mm'}$ (eq.~\ref{eqn:Gmm}) for spin 0,
 which has a more straightforward construction.

Our method can also be used in conjunction with the ubiquitous HEALPix grid.  Transforms by our method onto a sufficiently fine equiangular grid can be interpolated onto a HEALPix grid at high accuracy and a competitive speed.  Using this technique helps to combine the advantages of our approach (very high accuracy, fast supersampling and 
hence interpolation, and improved transform performance for multiple 
spin-$s$ transforms) with many elegant features of HEALPix, such as its hierarchical construction, and the equal area and near-regular shape of the pixels.


\section{Conclusions}  \label{sec:conclusions}
Borrowing the same trick that enables fast convolution on the sphere \citep{2001PhRvD..63l3002W}
and exploiting the relationship between Wigner $d$-functions and the
spin-$s$ spherical harmonics, we have presented a fast and exact ${\cal O}(L^3)$ algorithm which performs
spherical harmonic transforms for several spin $s$ functions with band-limit
$L$ using a single code.  
In particular, our 
method is computationally very  
efficient if several distinct spin transforms have to be computed because only one set of special
functions are needed.  
The algorithm's scaling with the number of pixels is subdominant; the extra cost for transforming a band-limited function from harmonic space to a high-resolution, oversampled map (or vice versa) scales modestly like a two-dimensional FFT.
%
%
This opens possibilities for efficient interpolation of spin-transformed fields at high resolution, which can be beneficial, for example, when simulating or analyzing gravitational lensing of the CMB.

\acknowledgments

Some of the results in this paper have been derived using the HEALPix  \citep{2005ApJ...622..759G} package.
KMH receives support from NASA-JPL subcontract 1363745.
BDW acknowledges support through NASA-JPL subcontracts 1236748 and 1371158 and NSF grant 07-08849 and thanks 
 Caltech and JPL for hospitality while part of this work was being completed.

\appendix

\section{Symmetries of Wigner $\Delta$-function}
\label{app:wigner_symm}
\begin{eqnarray}
  \Delta^l_{(-m')m} &=(-1)^{l+m} \Delta_{m'm} \quad &\mbox{(mirror first index)}\\ \nonumber
  \Delta^l_{m'(-m)} &=(-1)^{l+m'} \Delta_{m'm} \quad &\mbox{(mirror second index)}\\ \nonumber
  \Delta^l_{mm'} &=(-1)^{m'-m} \Delta_{m'm} \quad &\mbox{(swap indices)}.
\end{eqnarray}

\section{Exact quadrature of integral $I_{m'm}$}
 \label{quadrature}



\begin{figure}
\begin{center}
\includegraphics[width=\figwidth]{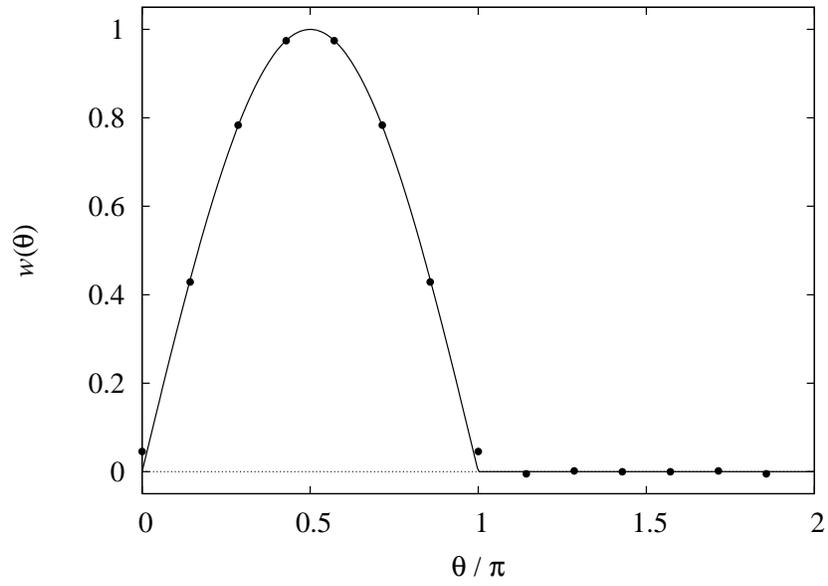}
\end{center}
\caption{Quadrature weights $w_r(\theta)$ when the number of samples of the extended function $F$ is $N'_\theta=14$ ($N_\theta = 8$), plotted against $\sin(\theta)$ for
  $\theta<\pi$, the weight function in the
  continuum limit.}
\label{fig:weights}
\end{figure}

We extend the function on the sphere in the $\theta$ direction via
\beq
F(\theta, \phi) = 
\left\{ \begin{array}{ll}
 f(\theta,\phi), & \theta \leq \pi \\
(-1)^s f(2\pi - \theta,\phi + \pi), & \theta > \pi
\end{array} \right.
\eeq
Because $F$ is the same as $f$ for $0 \leq \theta \leq \pi$, $F$ can take the place of $f$ in the integral for $I_{mm'}$ without changing its value.
This is convenient because $F$ (but not $f$) can be written as the band limited Fourier sum
\beq
F(\theta,\phi) = \sum_{kn=-L}^{L} F_{kn} \exp(i k \theta) \exp(i n \phi) 
\eeq
Substituting this into the expression for $I_{m'm}$ (Eq.~\ref{eqn:Imm}), we can rearrange to yield
\begin{equation}
 I_{m'm}= \sum_{kn=-L}^{L} F_{kn} \left[ \int_0^{2\pi} d\phi \exp \left( i(n-m)\phi \right) \right] \left[ \int_0^\pi d\theta \exp\left(i(k-m')\theta \right) \sin\theta \right] 
\end{equation}
The first integration, over $\phi$, is simply $2\pi\delta_{nm}$.  The second 
integration, over $\theta$, can be evaluated to yield a weight for the sum over the Fourier
coefficients. The result is
\beq
I_{m'm}=2\pi
\sum_{ k=-{L}}^{L}F_{km} w(k-m'),
\label{eqn:Impm_conv}
\eeq
where
\beq
w(p)= \int_0^\pi d\theta \exp(ip\theta) \sin\theta = \left\{ \begin{array}{ll}
0, & p\mbox{ even}\\
2/(1-p^{2}), & p\mbox{ odd}, p \neq \pm 1\\
\mp i\pi/2, & p = \pm 1
\end{array} \right.
\eeq
Note that (\ref{eqn:Impm_conv}) is written in the form of a discrete
convolution in Fourier space.  It therefore can be evaluated as an
multiplication in real space, so that $I_{m'm}$ is the discrete Fourier transform of $2 \pi w_r F$, 
 where 
\beq
w_r(q'\Delta\theta) = \sum_{p=-N'_\theta/2}^{N'_\theta/2-1} \exp (i p q' \Delta\theta) w(p)
\eeq
is the real-valued quadrature weight (see Fig.~\ref{fig:weights}) at the sampled $\theta$ values.  This gives weight to the pixels different than the $\sin\theta$ weight appearing in a naive Riemann sum, in particular giving weight to the poles which would have none in the naive scheme.  The notation $N'_\theta$ denotes the number of samples
in the function $F$ extended in $\theta$, which depending on the pixel
implementation, will be about twice the number of samples $N_\theta$ of $f$.  In our implementation $N'_\theta = 2 (N_\theta - 1)$.

Therefore we finally have
\beq
I_{m'm} = \frac{2\pi}{ N'_\theta N_\phi} \sum_{q'=0}^{N'_\theta-1} \sum_{q=0}^{N_\phi-1} \exp(-i m' q' \Delta\theta) \exp(-im q \Delta\phi) w_r(q' \Delta\theta) F(q' \Delta\theta, q \Delta\phi )
\eeq
where
\begin{equation}
  \Delta\theta = \frac{2\pi}{N'_\theta} \qquad \qquad  \Delta\phi = \frac{2\pi}{N_\phi} 
\end{equation}

To summarize, $I_{m'm}$ can be evaluated exactly by tabulating the weights
(via a one-dimensional FFT), multiplying the extended function by them,
then computing a two-dimensional FFT.  

%
%



\bibliography{fast_spin_s}
\bibliographystyle{apj}  

\end{document}